\documentclass[runningheads]{llncs}

\usepackage{eccv}

\usepackage{eccvabbrv}

\usepackage{graphicx}
\usepackage{booktabs}

\usepackage[accsupp]{axessibility}  

\usepackage{hyperref}

\usepackage{orcidlink}

\begin{document}


\title{ViMoWear: Visual Motion-Guided sEMG-IMU Representation Learning for Subject-Independent Thumb Gesture Recognition} 

\titlerunning{ViMoWear for Subject-Independent Thumb Gesture Recognition}

\author{Wenjuan Zhong\inst{}\orcidlink{0000-0001-6615-3809} \and
Chenfei Ma\inst{}\orcidlink{0000-0003-3136-0765} \and
Kianoush Nazarpour\inst{}\orcidlink{0000-0003-4217-0254
}}


\institute{University of Edinburgh, United Kingdom\\}

\maketitle

\begin{abstract}

Wearable sensing enables intuitive hand gesture recognition for human--computer interaction, augmented reality, and prosthetic control, yet subject--independent recognition remains challenging because wearable signals provide only indirect and highly subject-specific observations of hand motion. Although visual information can improve wearable gesture recognition, requiring it during inference increases sensing complexity and limits practical deployment. We propose ViMoWear, a visual-motion-guided framework that leverages synchronized 3D hand motion as training-only supervision while requiring only wearable sensing for gesture classification at inference. Specifically, Motion-Guided Cross-Subject Contrastive Learning (MGCL) promotes subject-robust representations, and Thumb-Aware Masked Motion Reconstruction (TMMR) preserves fine-grained motion information. The leave-one-subject-out experiments on a synchronized sEMG--IMU--pose dataset demonstrate consistent improvements over supervised baselines across multiple sensing configurations, while the learned representations also support classifier-free retrieval. The proposed training-only visual motion supervision improves the generalization of wearable representations to unseen subjects.

\end{abstract}

\section{Introduction}
\label{sec:intro}
Hand pose estimation from wearable sensing is a building block for augmented reality, prosthetic control, and physical intelligence \cite{RN1401, krasoulis2017improved, krasoulis2019multi, kaifosh2025generic, zhong2025deep}. Specifically, the thumb plays a central role in human hand function \cite{emerson1996anatomy,yue2017hand}, and thumb-oriented movements are also fundamental to modern human--computer interaction \cite{kin2024stmg,kieliba2021robotic,yang2025non}. Everyday interactions with smartphones, smartwatches, and virtual interfaces rely heavily on thumb gestures such as swiping, tapping, and pinching. Unlike traditional whole-hand postures \cite{pizzolato2017comparison, geng2016gesture, jiang2021open, zhong2023spatio}, these interactions are typically dynamic and transient, involving continuous movement rather than prolonged static poses. Accurately modelling these dynamic thumb movements from wearable sensing remains a significant challenge for intuitive and robust human--computer interaction.

This challenge largely stems from the nature of wearable sensing itself. Unlike vision, which directly captures hand motion, wearable sensors provide only indirect observations of the underlying movement. Surface electromyography (sEMG) records the electrical activity generated during muscle contractions, whereas inertial measurement units (IMUs) measure the resulting wrist kinematics. These measurements are highly sensitive to subject-specific factors, including anatomy, muscle recruitment strategies, sensor placement, and movement execution~\cite{cote2020unsupervised,campbell2020current, buskirk1970reproducibility, scheme2013training, kyranou2018causes,farina2014extraction}. Consequently, the same thumb gesture can produce substantially different wearable signals across individuals, while different gestures may exhibit similar transient signal patterns~\cite{zhong2026optimizing}. Learning representations that capture the underlying motion, rather than subject-specific signal characteristics, therefore remains difficult when relying solely on wearable data.

Visual observations provide a natural source of supervisory information for addressing this challenge. Three-dimensional (3D) hand pose directly captures the executed hand motion and is mostly independent of subject-specific physiological characteristics. However, existing multi-modal approaches typically fuse wearable and visual modalities during both training and inference, improving recognition performance at the cost of increased sensing complexity and reduced practicality for wearable deployment~\cite{xi2026egoemg, he2025gaze}. An alternative is to treat visual motion as training-only supervision that is available only during training to guide representation learning. In this paradigm, visual motion guides the learning of motion-centric wearable representations while preserving wearable-only classification \cite{cui2025cpep, cui2026embridge, gilardini2026kinembed}. However, this training paradigm has received little attention for subject-independent wearable gesture recognition.

We propose ViMoWear, a visual-motion-guided framework that leverages synchronized 3D hand motion as training-only supervision to learn generalizable wearable representations. The main contributions of this work are as follows:

\begin{itemize}

\item The ViMoWear is the first model to perform jointly pretraining from synchronized sEMG, IMU, and 3D hand motion.

\item We propose a pretraining objective that combines Motion-Guided Cross-Subject Contrastive Learning (MGCL) and Thumb-Aware Masked Motion Reconstruction (TMMR) to jointly learn subject-invariant and motion-aware wearable representations.

\item We conduct evaluations under the leave-one-subject-out (LOSO) protocol across sEMG-, IMU-, and sEMG+IMU-based thumb gesture recognition, demonstrating the effectiveness of ViMoWear for subject-independent thumb gesture recognition.
\end{itemize}

\begin{figure}[tb]
  \centering
  \includegraphics[width=1\textwidth]{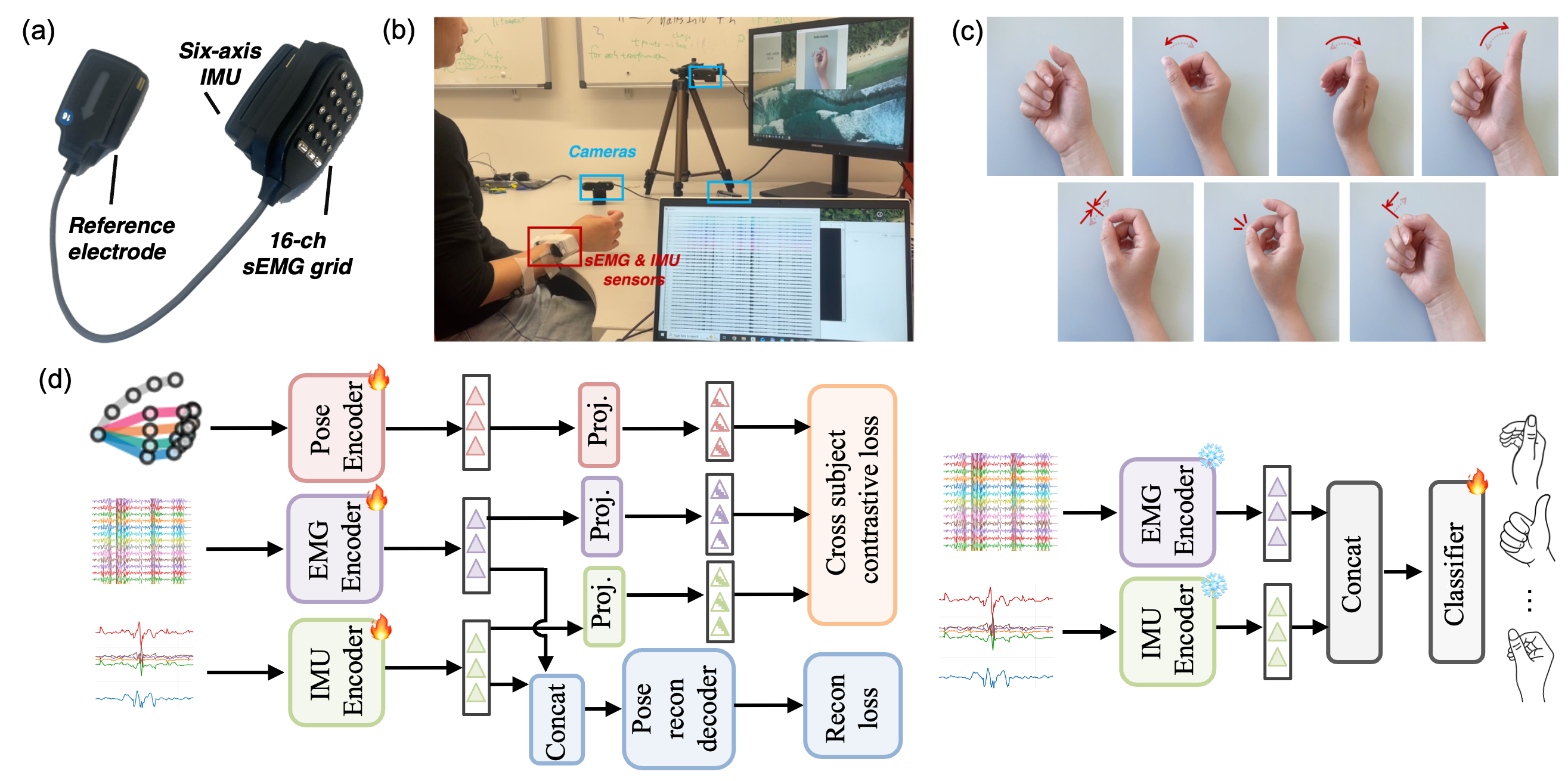}
   \caption{(a) Trigno sEMG and IMU sensors. (b) Experimental setup  for synchronized sEMG, IMU, and multi-camera acquisition. (c) The seven thumb gestures included in the dataset. (d) Overview of the proposed framework, consisting of first-stage visual-motion-guided wearable representation learning (left) and second-stage wearable-only gesture recognition (right). }
  \label{system}
\end{figure}

\section{Related Work}
Recent studies have explored using visual motion information to improve wearable representation learning beyond conventional supervised gesture classification. CPEP~\cite{cui2025cpep} and EMBridge~\cite{cui2026embridge} leverage the emg2pose dataset~\cite{salter2024emg2pose} to learn pose-informed sEMG representations through cross-modal pretraining, demonstrating improved downstream gesture recognition. Similarly, KinEMbed~\cite{gilardini2026kinembed} exploits hand kinematics to supervise sEMG representation learning for continuous joint-angle regression on the NinaPro DB8 dataset~\cite{krasoulis2019effect}. 

However, existing methods primarily focus on sEMG as the sole wearable modality, leaving the potential benefits of integrating complementary wearable sensors, such as IMUs, largely unexplored. Furthermore, the effectiveness of pose-informed wearable representation learning under strict subject-independent evaluation remains insufficiently studied. Although LOSO cross-validation is widely regarded as the most realistic evaluation protocol for practical wearable deployment, relatively few sEMG gesture recognition studies adopt this setting. Existing LOSO results on benchmark datasets such as NinaPro and CapgMyo typically report recognition accuracies around or below 50\%~\cite{du2017surface, islam2024surface}, illustrating the challenge of generalizing to unseen users.

\section{Method}
\subsection{Dataset Collection}
 We collected a synchronized tri-modal dataset comprising wrist-worn sEMG, IMU, and multi-view video recordings, as illustrated in Fig.~\ref{system}. The sEMG signals were acquired using two Trigno (Delsys Inc., USA) Maize $4\times4$ electrode arrays (5~mm spacing). IMU signals were recorded using two Trigno Avanti sensors, providing six-axis inertial measurements. Each IMU was mounted above a sEMG array using a custom 3D-printed holder (Fig.~\ref{system}(a)), forming two sensing modules placed on the extensor and flexor sides of the dominant wrist (Fig.~\ref{system}(b)). sEMG and IMU signals were sampled at 1000~Hz and 370~Hz, respectively, while synchronized multi-view videos were simultaneously recorded.

Thirty-three right-handed participants were recruited. After excluding two participants due to acquisition failures, the final dataset contained recordings from 31 participants. All participants provided written informed consent under an approved institutional ethics protocol.

\subsection{Preprocessing}
Seven thumb gestures were considered: \textit{rest},  \textit{swipe-left}, \textit{swipe-right}, \textit{swipe-up}, \textit{pinch-index}, \textit{hold-middle}, and \textit{tap} (Fig.~\ref{system}(c)). Each trial lasted 2~s.  Unlike conventional static gesture datasets, participants performed the gestures using a natural movement path, mimicking touchscreen or virtual-interface interactions. Consequently, the recorded signals primarily capture dynamic movement transitions. The experiment comprised ten blocks, each containing 37 trials: five repetitions of seven gesture classes presented in random order, plus initial and final \textit{rest} trials for IMU calibration. Participants rested for approximately 2~min between blocks. 

Raw sEMG signals were filtered using a fourth-order Butterworth band-pass filter (20-450~Hz). The IMU signals were calibrated using the first and last rest trials and filtered using fourth-order Butterworth low-pass filters with 10~Hz (acceleration) and 20~Hz (gyroscope). The recorded videos were processed using a pre-trained neural network from the MediaPipe framework~\cite{lugaresi2019mediapipe} to extract 2D coordinates of 21 joint landmarks of the hand. The key points were then triangulated into 3D coordinates using the Anipose library \cite{karashchuk2021anipose}.

To obtain gesture-centred motion segments, movement onset and offset were estimated from the displacement, velocity, and acceleration of the relative thumb trajectory. For \textit{rest} trials, a fixed temporal window was adopted. Finally, the reconstructed hand pose (21 landmarks, 63 dimensions) was transformed into a wrist-relative coordinate system by subtracting the wrist landmark from the remaining 20 landmarks, resulting in a 20-landmark, 60-dimensional 3D pose representation.

\subsection{Problem Definition}
Our objective is to learn wearable signal representations that benefit from visual motion supervision during training while requiring only wearable sensing during downstream gesture classification. Given synchronized sEMG, IMU, and 3D hand pose sequences, the goal is to optimize a wearable encoder that generalizes across unseen subjects without relying on visual information during inference. 

Formally, Let $\mathcal{D}=\{(\mathbf{x}^{e}_{i}, \mathbf{x}^{m}_{i}, \mathbf{x}^{p}_{i}, y_i, s_i)\}_{i=1}^{N}$ denote the synchronized multi-modal dataset, where $N$ is the total number of valid motion segments. For the $i$-th trial, $\mathbf{x}^{e}_{i}\in\mathbb{R}^{C_e \times T}$ denotes the sEMG sequence, $\mathbf{x}^{m}_{i}\in\mathbb{R}^{C_m \times T}$ denotes the IMU sequence, and $\mathbf{x}^{p}_{i}\in\mathbb{R}^{C_p \times T}$ denotes the 3D hand pose sequence. In our dataset, $C_e=32$ corresponds to two $4\times4$ sEMG arrays, $C_m=12$ corresponds to two six-axis IMU sensors, and $C_p=60$ corresponds to the wrist-relative 3D coordinates of 20 hand landmarks. All modalities are temporally aligned and resampled to $T=700$ time samples. The gesture label is denoted by $y_i\in\mathcal{Y}$, where $\mathcal{Y}$ contains $K=7$ thumb gesture classes, and $s_i$ denotes the subject identity.

\section{Architecture}
The architecture employs a two-stage training process (Fig.~\ref{system}(d)). In the first stage, ViMoWear is pretrained to learn semantically structured and subject-robust wearable representations. In the second stage, depending on the downstream modality setting, the model uses sEMG alone, IMU alone, or both with their corresponding encoders, to train a lightweight classifier.

\subsection{ViMoWear}
\subsubsection{Wearable and Motion Encoders}
ViMoWear contains separate encoders for sEMG, IMU, and 3D hand pose. The sEMG encoder $f_e(\cdot)$ maps an input sequence $\mathbf{x}^{e}_{i}\in\mathbb{R}^{C_e\times T}$ into a latent embedding $\mathbf{z}^{e}_{i}\in\mathbb{R}^{d}$. Similarly, the IMU encoder $f_m(\cdot)$ maps $\mathbf{x}^{m}_{i}\in\mathbb{R}^{C_m\times T}$ to $\mathbf{z}^{m}_{i}\in\mathbb{R}^{d}$, and the pose encoder $f_p(\cdot)$ maps $\mathbf{x}^{p}_{i}\in\mathbb{R}^{C_p\times T}$ to $\mathbf{z}^{p}_{i}\in\mathbb{R}^{d}$:
\[
\mathbf{z}^{e}_{i}=f_e(\mathbf{x}^{e}_{i}), \quad
\mathbf{z}^{m}_{i}=f_m(\mathbf{x}^{m}_{i}), \quad
\mathbf{z}^{p}_{i}=f_p(\mathbf{x}^{p}_{i}).
\]

The sEMG, IMU, and pose encoders share a three-block temporal CNN architecture~\cite{meng2022user,zhong2026optimizing}, differing in their input dimensionality. The convolutional blocks use channel dimensions $[64,128,128]$ with kernel sizes $[7,5,3]$, batch normalization, and ReLU activations, with temporal pooling applied after the first two blocks. Adaptive average pooling and a two-layer MLP then produce a $128$-dimensional trial-level representation.

\subsubsection{Motion-Guided Cross-Subject Contrastive Learning}
Hand pose provides structured kinematic information about gesture execution. MGCL exploits this information by using pose representations as motion anchors for learning wearable embeddings.

Modality-specific projection heads map the encoder outputs into a shared contrastive space:
\[
\mathbf{q}^{e}_{i}=h_e(\mathbf{z}^{e}_{i}), \quad
\mathbf{q}^{m}_{i}=h_m(\mathbf{z}^{m}_{i}), \quad
\mathbf{q}^{p}_{i}=h_p(\mathbf{z}^{p}_{i}),
\]
where $h_e(\cdot)$, $h_m(\cdot)$, and $h_p(\cdot)$ denote two-layer nonlinear projection heads for sEMG, IMU, and pose, respectively, with hidden and output dimensions of 128. A ReLU activation is applied between the two linear layers, followed by $\ell_2$ normalization of the projected embeddings.

Given a mini-batch $\mathcal{B}$, the positive pose set for each wearable anchor $i$ comprises samples of the same gesture from different subjects:
\[
\mathcal{P}(i)
=
\left\{
j\in\mathcal{B}:
y_j=y_i,\;
s_j\neq s_i
\right\}.
\]

For a wearable modality $a\in\{e,m\}$, the cross-subject contrastive loss is
\[
\mathcal{L}_{a,p}
=
-\frac{1}{|\mathcal{B}|}
\sum_{i\in\mathcal{B}}
\frac{1}{|\mathcal{P}(i)|}
\sum_{j\in\mathcal{P}(i)}
\log
\frac{
\exp\left(
\mathrm{sim}(\mathbf{q}^{a}_{i},\mathbf{q}^{p}_{j})/\tau
\right)
}{
\displaystyle
\sum_{k\in\mathcal{B}}
\exp\left(
\mathrm{sim}(\mathbf{q}^{a}_{i},\mathbf{q}^{p}_{k})/\tau
\right)
},
\]

where $\mathrm{sim}(\cdot,\cdot)$ denotes cosine similarity and $\tau$ is the temperature parameter. The denominator spans all pose embeddings in the mini-batch; hence, same-subject same-label samples are excluded from the positive set but remain as competing targets. When no cross-subject positive is available, the paired pose sample is used instead.

For single-modality pretraining, MGCL aligns either sEMG or IMU with pose:
\[
\mathcal{L}_{\mathrm{MGCL}}=\mathcal{L}_{e,p}
\quad \text{or} \quad
\mathcal{L}_{\mathrm{MGCL}}=\mathcal{L}_{m,p}.
\]
For the sEMG+IMU setting, sEMG and IMU are aligned with pose separately:
\[
\mathcal{L}_{\mathrm{MGCL}}
=
\mathcal{L}_{e,p}
+
\mathcal{L}_{m,p}.
\]

Similar to CLIP-style contrastive learning~\cite{radford2021learning}, the modality-specific encoders are trained jointly. Unlike  CPEP~\cite{cui2025cpep} and EMBridge~\cite{cui2026embridge}, however, MGCL does not use a standard instance-level InfoNCE loss (Instance-NCE), where only the paired wearable--pose sample is treated as positive. Instead, positives are defined as pose samples with the same gesture label from different subjects. This changes the objective from trial-level correspondence learning to cross-subject gesture-level alignment.

\subsubsection{Thumb-Aware Masked Motion Reconstruction}

Beyond the contrastive loss MGCL, we add TMMR objective to enrich the representation learning process. Let
$\mathbf{M}_{i}\in\{0,1\}^{C_p\times T}$
denote a randomly generated binary temporal mask, where entries equal to one indicate the pose coordinates selected for reconstruction. In the sEMG+IMU setting, the reconstruction decoder, a two-layer MLP module, predicts the complete pose sequence from the fused wearable representation:
\[
\hat{\mathbf{x}}^{p}_{i}
=
r([\mathbf{z}^{e}_{i};\mathbf{z}^{m}_{i}]).
\]
For single wearable modality variants, the decoder uses the corresponding wearable representation, e.g.,
$r(\mathbf{z}^{e}_{i})$ or $r(\mathbf{z}^{m}_{i})$. Because the recognition task focuses on thumb motion, different hand landmarks do not contribute equally. TMMR applies a landmark-wise weighting matrix
$\mathbf{W}\in\mathbb{R}^{C_p\times T}$
that assigns greater importance to thumb-related coordinates while preserving weaker supervision from the remaining hand landmarks. The reconstruction objective is
\[
\mathcal{L}_{\mathrm{TMMR}}
=
\frac{
\displaystyle
\sum_{i\in\mathcal{B}}
\left\|
\mathbf{W}\odot\mathbf{M}_{i}\odot
\left(
\hat{\mathbf{x}}^{p}_{i}-\mathbf{x}^{p}_{i}
\right)
\right\|^{2}_{2}
}{
\displaystyle
\sum_{i\in\mathcal{B}}
\left\|
\mathbf{W}\odot\mathbf{M}_{i}
\right\|_{1}
}.
\]
This formulation prioritizes the motion most relevant to the task while retaining contextual information from the overall hand configuration.

The complete ViMoWear pretraining objective is
\[
\mathcal{L}_{\mathrm{pre}}
=
\mathcal{L}_{\mathrm{MGCL}}
+
\lambda_{\mathrm{TMMR}}
\mathcal{L}_{\mathrm{TMMR}},
\]
with weight $\lambda_{\mathrm{TMMR}}$ >0.

\subsection{Wearable-Only Downstream Training}

After the first-stage pretraining, the pose encoder, pose projector, and reconstruction decoder are discarded. During the second stage, the pretrained wearable encoders are frozen, and gesture recognition is performed using linear probing (LP), where only the classifier $g_{\phi}(\cdot)$ is optimized.

For sEMG-only and IMU-only evaluation, the classifier is applied to the corresponding wearable embedding:
\[
\hat{y}_{i}
=
g_{\phi}(\mathbf{z}^{e}_{i})
\quad \text{or} \quad
\hat{y}_{i}
=
g_{\phi}(\mathbf{z}^{m}_{i}).
\]
For the sEMG+IMU setting, the two wearable embeddings are fused after encoding:
\[
\hat{y}_{i}
=
g_{\phi}([\mathbf{z}^{e}_{i};\mathbf{z}^{m}_{i}]).
\]
The downstream classification objective is
\[
\mathcal{L}_{\mathrm{cls}}
=
-\frac{1}{|\mathcal{B}|}
\sum_{i\in\mathcal{B}}
\log
p_{\theta,\phi}
\left(
y_i\mid\mathbf{x}^{w}_{i}
\right).
\]

\section{Experimental Setup}

\subsection{Subject-Independent Evaluation}
We evaluate subject-independent gesture recognition using a LOSO protocol. Three participants are reserved for validation, while each of the remaining 28 participants serves once as the unseen test subject. In every fold, the training, validation, and test participants are mutually exclusive. Consequently, all reported results are averaged over 28 test folds.

\subsection{Pretraining Configuration}
During first-stage pretraining, synchronized wearable and 3D pose sequences are used to learn ViMoWear representations. Depending on the downstream sensing configuration, an sEMG model, an IMU model, or a fused sEMG+IMU model is pretrained using the corresponding pose sequences as visual motion supervision by optimizing $\mathcal{L}_{\mathrm{pre}}$.

In the main experiments, the contrastive temperature was set to $\tau=0.07$, the pose masking ratio is set to $0.7$, and $\lambda_{\mathrm{TMMR}}=0.5$. For the landmark-wise weighting matrix $\mathbf{W}\in\mathbb{R}^{C_p\times T}$ in $\mathcal{L}_{\mathrm{TMMR}}$, thumb landmarks are assigned a reconstruction weight of $1.0$, whereas non-thumb landmarks are assigned a weight of $0.2$. Additionally, we evaluate two alternative reconstruction objectives for ablation. In $\mathcal{L}_{\mathrm{full-hand MMR}}$, all landmarks are assigned a reconstruction weight of $1.0$. In  $\mathcal{L}_{\mathrm{thumb-only MMR}}$, thumb landmarks are assigned a weight of $1.0$, whereas non-thumb landmarks are assigned a weight of $0$.

\subsection{Downstream Protocols}
Two evaluation protocols are considered. For LP, the pretrained wearable encoder is frozen, and a linear classifier is trained using gesture labels. For classifier-free retrieval (CFR), no downstream classifier is trained. Instead, each test wearable embedding retrieves the most similar pose embedding from the training set using cosine similarity, and the corresponding gesture label is assigned to the query. The supervised baselines train wearable encoders and the linear classifiers from scratch using gesture labels. Supervised baselines do not support CFR(nearest neighbor) classification in the embedding space. Performance is reported using balanced accuracy and macro-F1 score to account for possible class imbalance across gesture classes.

\subsection{Implementation Details}
The model was implemented in PyTorch and trained on four NVIDIA RTX 3090 GPUs. Adam was used for both pretraining and downstream optimization with a learning rate of $10^{-3}$, weight decay of $10^{-4}$, and a batch size of 512. Pretraining and downstream training were run for up to 200 and 100 epochs, respectively, with early stopping patience of 30 and 20 epochs. Unless otherwise stated, all experiments used a random seed of 42.

\section{Results and Discussion}

\subsection{Performance of Thumb Gesture Recognition}
Table~\ref{tab:modality} summarises the performance of ViMoWear under the LOSO protocol across sEMG-, IMU-, and fused wearable sensing configurations. Visual-motion-guided pretraining yields higher mean LP performance than the corresponding supervised model for all three sensing configurations. The main comparison is observed with fused sEMG+IMU, where ViMoWear improves balanced
accuracy from $73.22\%$ to $77.67\%$ and macro-F1 from $71.74\%$ to $76.43\%$. These correspond to gains of $4.45\%$ in balanced accuracy (Wilcoxon $p=0.0056$) and $4.70\%$in macro-F1 (Wilcoxon $p=0.0095$).

Fusing sEMG and IMU provides a further benefit. With ViMoWear, the fused representation reaches $77.67\%$ balanced accuracy, compared with $71.79\%$ for sEMG and $60.87\%$ for IMU, with significant paired improvements over both single-modality configurations (Wilcoxon $p<0.001$). Their combination therefore produces a more complete representation of hand motion than either modality alone. Importantly, the performance improvements obtained with ViMoWear are consistently observed across all sensing configurations, indicating that the proposed pretraining strategy is complementary to the sensing modality itself rather than relying on a particular sensor type.

\begin{figure}[tb]
  \centering
  \includegraphics[width=1\textwidth]{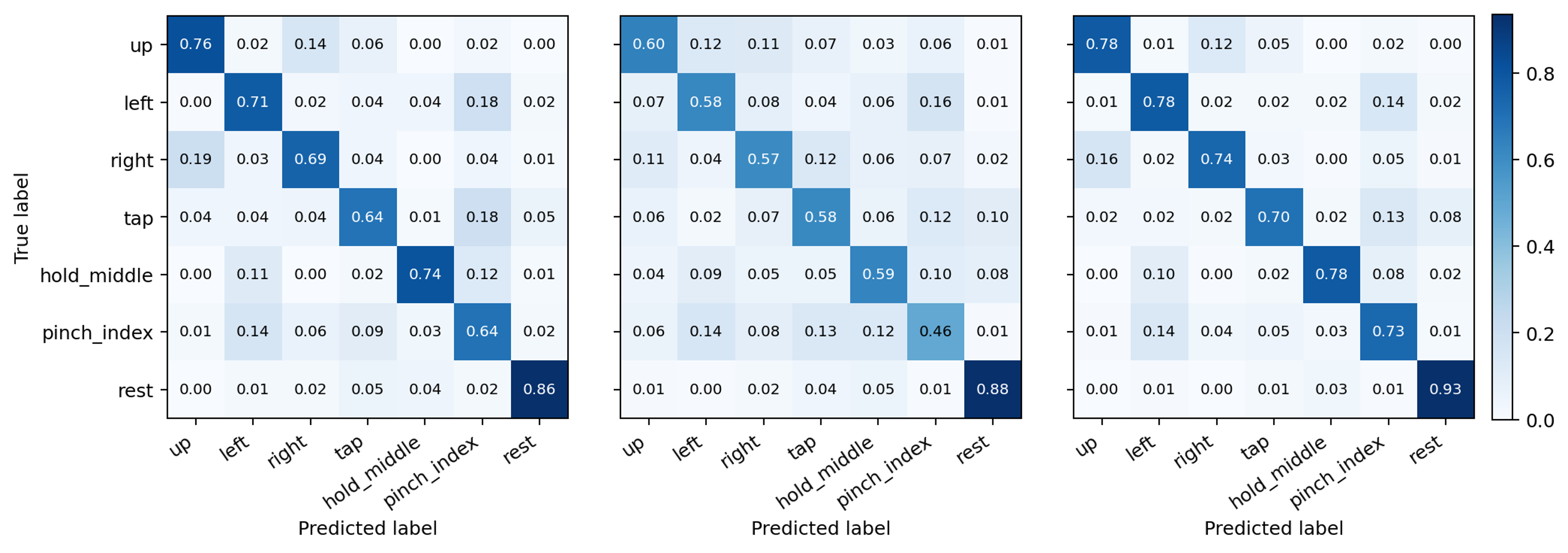}
   \caption{Aggregated confusion matrices for linear probing on unseen participants. Left ViMoWear with sEMG; Middle:ViMoWear with IMU;  Right: ViMoWear with fused sEMG and IMU. }
  \label{fig:cm}
\end{figure}

Figure~\ref{fig:cm} provides a class-wise view of the improvements. It shows that the performance gains of EMG+IMU modality are not confined to specific gesture classes. Instead, ViMoWear consistently improves the recognition of most thumb gestures, while the remaining errors are primarily limited to gestures with intrinsically similar movement patterns, such as \textit{up} versus \textit{right} and \textit{left} versus \textit{pinch-index}. 

\begin{table}[tb]
  \caption{Effect of wearable input modality and visual motion supervision. Supervised models are trained from scratch. ViMoWear uses 3D pose only during pretraining and performs wearable-only gesture classification. LP denotes linear probing and CFR denotes classifier-free retrieval.}
  \label{tab:modality}
  \centering
  \begin{tabular}{@{}lcccc@{}}
    \toprule
    Method & \multicolumn{2}{c}{Balanced Acc. (\%)} &
    \multicolumn{2}{c}{Macro-F1 (\%)}\\
    \cmidrule(lr){2-3}\cmidrule(lr){4-5}
     & LP & CFR & LP & CFR\\
    \midrule
    Supervised sEMG
    & $69.71 \pm 10.73$ & --
    & $67.73 \pm 11.67$ & --\\
    Supervised IMU
    & $58.78 \pm 10.20$ & --
    & $56.97 \pm 10.38$ & --\\
    Supervised sEMG+IMU
    & $73.22 \pm 10.59$ & --
    & $71.74 \pm 11.60$ & --\\
    \midrule
    ViMoWear (sEMG)
    & $71.79 \pm 10.56$ & $70.76 \pm 12.07$
    & $70.08 \pm 11.45$ & $69.08 \pm 12.95$\\
    ViMoWear (IMU)
    & $60.87 \pm 10.69$ & $58.74 \pm 12.19$
    & $59.43 \pm 10.79$ & $57.46 \pm 12.44$\\
    ViMoWear (sEMG+IMU)
    & $\mathbf{77.67 \pm 8.70}$ & $\mathbf{73.60 \pm 8.40}$
    & $\mathbf{76.43 \pm 9.66}$ & $\mathbf{72.41 \pm 9.22}$\\
    \bottomrule
  \end{tabular}
\end{table}

\subsection{Effect of the Pretraining Objectives}

Table~\ref{tab:loss} investigates the contribution of each pretraining objective using fused sEMG and IMU. Replacing conventional Instance-NCE with MGCL substantially improves both LP and CFR performance, increasing LP balanced accuracy from $69.05\%$ to
$76.94\%$ and macro-F1 from $67.15\%$ to $75.65\%$ (both $p<0.001$).
This highlights the importance of cross-subject, gesture-level alignment for learning representations that generalize across unseen participants.

Adding motion reconstruction provides only modest gains beyond MGCL under LP Full-hand MMR and TMMR achieve nearly identical balanced accuracy ($77.68\%$ and $77.67\%$), with neither significantly outperforming MGCL alone. The differences among reconstruction strategies become more apparent under CFR: TMMR achieves the highest mean performance, although its advantage over MGCL and full-hand MMR is not statistically significant. In contrast, restricting reconstruction to thumb landmarks reduces CFR performance, with TMMR significantly outperforming thumb-only MMR ($p=0.0110$). Together, these results suggest that the benefit of motion reconstruction lies not in isolating the thumb, but in emphasizing task-relevant thumb motion while preserving the broader hand configuration as motion context.
\begin{table}[tb]
  \caption{Ablation of the ViMoWear pretraining objectives using fused sEMG and IMU. All models use wearable-only downstream classification.}
  \label{tab:loss}
  \centering
  \begin{tabular}{@{}lcccc@{}}
    \toprule
    Pretraining objective &
    \multicolumn{2}{c}{Balanced Acc. (\%)} &
    \multicolumn{2}{c}{Macro-F1 (\%)}\\
    \cmidrule(lr){2-3}\cmidrule(lr){4-5}
     & LP & CFR & LP & CFR\\
    \midrule
    Instance-NCE
    & $69.05 \pm 9.18$
    & $60.60 \pm 9.87$
    & $67.15 \pm 10.58$
    & $59.09 \pm 10.53$\\
    
    MGCL
    & $76.94 \pm 8.14$
    & $71.89 \pm 9.77$
    & $75.65 \pm 8.97$
    & $70.52 \pm 10.66$\\
    
    MGCL + thumb-only MMR
    & $77.24 \pm 8.49$
    & $70.05 \pm 7.62$
    & $76.09 \pm 9.43$
    & $68.79 \pm 8.19$\\

    MGCL + full-hand MMR
    & $\mathbf{77.68 \pm 9.30}$
    & $72.39 \pm 9.30$
    & $\mathbf{76.47 \pm 10.16}$
    & $71.39 \pm 10.04$\\

    MGCL + TMMR
    & $77.67 \pm 8.70$
    & $\mathbf{73.60 \pm 8.40}$
    & $76.43 \pm 9.66$
    & $\mathbf{72.41 \pm 9.22}$\\
    
    \bottomrule
  \end{tabular}
\end{table}

\subsection{Freeze Backbone for Downstream}
Table~\ref{tab:linear-vs-finetune} compares frozen LP with full fine-tuning (FT) after ViMoWear pretraining. In full FT, both the wearable encoder and classifier are updated. Under the current optimization setting, LP achieves comparable or slightly higher balanced accuracy than FT across all sensing configurations. For sEMG+IMU, LP achieves $77.67\%$ balanced accuracy compared with $76.45\%$ for FT. These results suggest that the pretrained representations remain effective without end-to-end adaptation.

\begin{table}[tb]
  \caption{Comparison of balanced accuracy (\%) between linear probing and full fine-tuning after ViMoWear pretraining. Positive values of $\Delta$ indicate an advantage for freezing the pretrained wearable encoder.}
  \label{tab:linear-vs-finetune}
  \centering
  \begin{tabular}{@{}lccc@{}}
    \toprule
    Wearable input &
    Freeze  &
    Fine-tune &
    $\Delta$ Freeze--FT\\
    \midrule
    sEMG
    & $\mathbf{71.79 \pm 10.56}$
    & $70.74 \pm 10.86$
    & $+1.05$\\
    IMU
    & $\mathbf{60.87 \pm 10.69}$
    & $58.95 \pm 10.64$
    & $+1.92$\\
    sEMG+IMU
    & $\mathbf{77.67 \pm 8.70}$
    & $76.45 \pm 10.34$
    & $+1.22$\\
    \bottomrule
  \end{tabular}
\end{table}

\section{Broader Impacts and Limitations}
ViMoWear highlights an alternative paradigm for multimodal wearable learning, in which visual motion is exploited exclusively as privileged supervision during training rather than as an additional sensing modality during inference. This decouples representation learning from inference-time sensing requirements, enabling rich motion information to improve wearable representations while preserving the simplicity, portability, and low cost of wearable-only systems. The same principle may extend beyond thumb gesture recognition to other wearable sensing problems where high-quality supervisory modalities, such as vision or motion capture, are available during data collection but impractical in real-world use.

Several limitations should also be acknowledged. First, ViMoWear requires synchronized wearable and 3D motion recordings during pretraining, making the quality of representation learning dependent on paired multimodal datasets. Second, the current evaluation uses vision-derived thumb trajectories to define gesture boundaries. A fully wearable pipeline would require gesture segmentation directly from sEMG and IMU. Third, the current evaluation is limited to a single thumb gesture dataset collected with one wearable configuration. Further validation on more diverse hand activities, sensor layouts, and larger-scale datasets is needed to assess the generality of the proposed framework.

\section{Conclusion}
ViMoWear suggests that visual motion can improve subject-independent wearable gesture recognition without necessarily being required during inference. Instead, synchronized 3D hand motion can serve as privileged supervision during pretraining to learn wearable EMG-IMU representations that generalize effectively across unseen subjects. These findings suggest a practical paradigm that combines rich visual supervision during training with lightweight wearable sensing during deployment, providing a promising direction for future wearable human--computer interaction systems.


%
%
\bibliographystyle{splncs04}
\bibliography{main}
\end{document}